%% file: main.tex
\documentclass[conference]{IEEEtran}
\usepackage{amsmath}
\usepackage{graphicx}
\usepackage{textcomp}
\usepackage{xcolor}
\usepackage[linesnumbered,ruled,vlined]{algorithm2e}
\usepackage{algorithmicx}
\usepackage{algpseudocode}
\usepackage{pdfpages}
\usepackage{pifont}
\usepackage{url}

\usepackage{soul}
\usepackage{acro}
\usepackage{enumitem}

\usepackage{multirow}

\usepackage{booktabs}
\usepackage{tikz}

\usetikzlibrary{arrows,calc,patterns,positioning,shapes,shapes.symbols}

\usepackage{cleveref}
\crefname{section}{Section}{Sections}
\crefname{appendix}{Section}{Sections}
\crefname{figure}{Fig.}{Figs.}
\Crefname{figure}{Figure}{Figures}
\crefname{table}{Table}{Tables}
\crefname{equation}{Eq.}{Eqs.}
\Crefname{equation}{Equation}{Equations}
\crefname{algorithm}{Alg.}{Algs.}
\Crefname{algorithm}{Algorithm}{Algorithms}

\input{acronyms}
\begin{document}

\title{MeMoir: A Software-Driven Covert Channel based on Memory Usage}

\author{\IEEEauthorblockN{Jeferson Gonz\'alez-G\'omez\IEEEauthorrefmark{1}\IEEEauthorrefmark{2}, Jose Alejandro Ibarra-Campos\IEEEauthorrefmark{3}, Jesus Yamir Sandoval-Morales\IEEEauthorrefmark{4},\\ Lars Bauer, J\"{o}rg Henkel\IEEEauthorrefmark{2}}

\IEEEauthorblockA{\IEEEauthorrefmark{1} \textit{Instituto Tecnol\'ogico de Costa Rica (TEC)}}
\IEEEauthorblockA{\IEEEauthorrefmark{2} \textit{University of Exeter}}
\IEEEauthorblockA{\IEEEauthorrefmark{3} \textit{Georgia Institute of Technology}}
\IEEEauthorblockA{\IEEEauthorrefmark{3}
\textit{Karlsruhe Institute of Technology (KIT), Chair for Embedded Systems (CES)} \\
\textit{Corresponding author:} Jeferson Gonz\'alez-G\'omez, jeferson.gonzalez@kit.edu}}

\maketitle

\begin{abstract}
Covert channel attacks have been continuously studied as severe threats to modern computing systems.
Software-based covert channels are a typically hard-to-detect branch of these attacks, since they leverage virtual resources to establish illegitimate communication between malicious actors.
In this work, we present MeMoir: a novel software-driven covert channel that, for the first time, utilizes memory usage as the medium for the channel.
We implemented the new covert channel on two real-world platforms with different architectures: a general-purpose Intel x86-64-based desktop computer and an ARM64-based embedded system.
Our results show that our new architecture- and hardware-agnostic covert channel is effective and achieves moderate transmission rates with very low error.
Moreover, we present a real use-case for our attack where we were able to communicate information from a Hyper-V virtualized enviroment to a Windows 11 host system.
In addition, we implement a machine learning-based detector that can predict whether an attack is present in the system with an accuracy of more than $95$\% with low false positive and false negative rates by monitoring the use of system memory.
Finally, we introduce a noise-based countermeasure that effectively mitigates the attack while inducing a low power overhead in the system compared to other normal applications.
\end{abstract}

\IEEEpeerreviewmaketitle

\section{Introduction} \label{sec:intro}

Covert channels are hidden means of communication that are used to secretly send data between two modules in a computer system or network. 
They take advantage of vulnerabilities in hardware and software to create a medium that allows information transfer and is difficult to detect by regular users~\cite{salwancovertchannels2013}.
Although these channels may be used for legitimate purposes, such as protecting confidential data, they can serve ill-intentioned attackers to spy on or spread malicious applications~\cite{steganography2020}. 

Several types of covert channels have been implemented in the literature, with notable recent works on thermal covert channels~\cite{HuangDetection2022, HuangOnCounter2020, gonzalezgpu2023} for hardware-supported attacks, and OS synchronization mechanisms~\cite{ZhangMex2023, shen2022mesattacks, EfanovPort2018} for those driven by software. 
With an increase in covert channel research in recent years, covert channels have become a threat to emerging computing systems~\cite{MiketicEmerging}.

   \begin{figure} [ht]
        \centering
        \includegraphics[width=\linewidth]{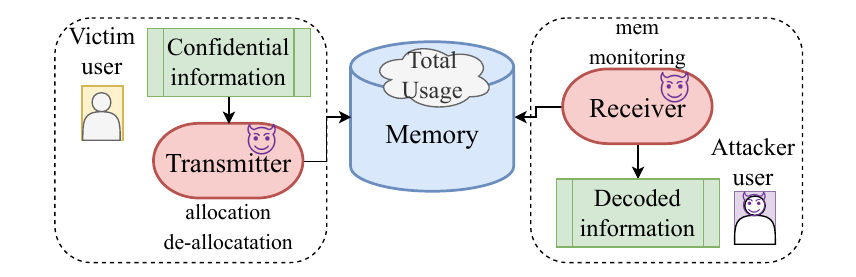}
        \caption{Overview of the new software-controlled memory-usage-based covert channel}
        \label{fig:overview}
    \end{figure}

In order to unveil a new type of covert channel, this paper deals with the implementation, detection, and mitigation of memory-usage-based threats.
To introduce our new software-based covert channel, \cref{fig:overview} shows an overview of the mechanism and actors involved in the attack form the perspective of a multi-tenant server..
In such an attack, the malicious transmitter application, which executes in a target victim user (private) context, has gained access to the unaware victim's confidential information, and it seeks to communicate the secret to other users in the system avoiding obvious direct mechanisms (e.g., shared memory, files, sockets, etc.) which are normally monitored \cite{MishraTime2023} and hence easy to detect.
In order to leak the secrets in a stealthy manner, as part of our new attack, the transmitter modulates the memory usage in the system, creating periodic patterns or memory allocations and de-allocations to encode the `1's and `0's of the message.
Under a second user's context (or any other nonvictim zone), a second malicious application ---the receiver--- reads the system's memory usage and decodes the message being sent.
Because the modulated signal (memory usage) is a virtual resource, it does not necessarily have a physical effect on the system which makes it hard to detect, especially if its existence is not yet unveiled.

\subsection*{Contributions}
In this work, we present the following contributions.
\begin{enumerate}
    \item We propose a novel software-driven covert channel that leverages the memory utilization of an application to effectively transmit information in a stealthy manner even under background noise memory usage.
    \item We present a real use case of our attack, where information is successfully communicated from an isolated virtual environment to a host system.
    \item  We develop a machine-learning-based detection technique for the novel covert channel attack while exploring the performance of different models and approaches from this domain.
    \item We implement a countermeasure for this new attack while discussing the implications of such techniques both in the attack's effectiveness and the system's performance.
\end{enumerate}

\section{Related Work}

The implementation of covert channels is a well-researched topic in the literature. Various attacks using covert channels that exploit the system's physical resources have been documented, employing mediums such as CPU/GPU temperature~\cite{masti2015, Dhananjay2022, gonzalezgpu2023}, power in FPGA-based systems~\cite{Gnad2021, MathieuPower2023}, electromagnetic emissions from USB ports~\cite{Guri2016}, and timing differences in shared caches for virtual machines and cloud environments~\cite{MauriceC52015, Xu2011}. Although these hardware-based covert channels are effective, they are generally easier to detect and mitigate in both design and execution due to their physical nature~\cite{szefer2019survey}.

In contrast, software-based covert channels use virtual resources to facilitate communication between malicious applications, making detection more challenging~\cite{shen2022mesattacks}. Examples of such channels include OS-level synchronization mechanisms~\cite{ZhangMex2023, shen2022mesattacks}, page caches~\cite{GrussPage2019}, and TCP sockets for interprocess communication~\cite{EfanovPort2018}, among others.

Several studies have addressed memory-based covert channels. In~\cite{chen2019}, the authors estimate the current temperature of a DRAM module using cell decay rates, implementing a thermal covert channel with reliability up to 95\%. Saileshwar \emph{et al}.~\cite{saileshwar2021} exploits memory contention in shared resource systems, coordinating memory operation timings between transmitter and receiver applications to achieve a transfer speed of $1801$kB/s. To our knowledge, \textbf{memory usage-based covert channels have not been disclosed in the literature prior to this work}.

In terms of detection, notable research by Elsadig \& Gafar~\cite{elsadiggafar2022} reviews various Machine Learning (ML) techniques, such as neural networks and support vector machines, achieving high detection accuracy. Other classification techniques have also achieved high accuracy (more than 95\%) in practical tests for thermal covert channels~\cite{wang2023}, showing significant improvements over traditional threshold-based detectors for complex transmission schemes.

For protection strategies, Deustch \emph{et al.,}~\cite{deutsch2022} explores mitigating covert channels based on memory access patterns by defining a secret and independent code for memory operations. The proposed countermeasure profiles the transmitter application to build an optimized defense module based on memory access time patterns for side-channel attacks. Other countermeasure approaches include thermal noise generation~\cite{Rahimi2022noise} and resource management-based solutions such as Dynamic Voltage and Frequency Scaling (DVFS)\cite{HuangDetection2022, gonzalezgpu2023} and task migration\cite{Wu2021Task, GonzalezCache2023}.

\section{MeMoir: A Novel Software-Based Covert Channel via Memory Usage}
\subsection{Threat model}

The threat model for our attack draws inspiration from the existing covert channels discussed in the literature \cite{MishraTime2023, shen2022mesattacks}. Similarly to these channels, our model assumes the presence of two distinct applications: a transmitter (malware) and a receiver (spy), operating under different permission levels within a multitenant system. In this environment, users share the same hardware and operating system infrastructure, yet conventional communication methods such as file sharing and sockets are readily detectable due to inherent security measures such as permission restrictions.

The attack leverages a covert channel mechanism to facilitate stealthy communication between the transmitter and the receiver. The transmitter operates within the context of a victim user, exploiting access to private information without requiring elevated privileges. This is achieved by modulating the usage of memory to leak data to other users on the system. Similarly, the receiver operates within the context of an attacker user, decoding transmitted data by monitoring system memory usage through the operating system, also without needing elevated permissions.

In practical scenarios, such a transmitter malware could be distributed through malicious vendors or exploited via vulnerabilities in legitimate applications. An illustrative example is the recent vulnerability of xz utils \cite{cvexz, lins2024critical}, where a supply chain attack introduced a backdoor into the xz utils library. This incident almost resulted in the integration of malicious updates into major Linux distributions before being detected.

\subsection{Attack implementation}
\subsubsection{\textbf{Transmitter}}
This module's role is to encode and send the target data through the covert channel. 
In this context, it is seen as a malicious application with access to confidential information related to private data.

\begin{figure} [ht]
  \centering
  \includegraphics[width=\linewidth]{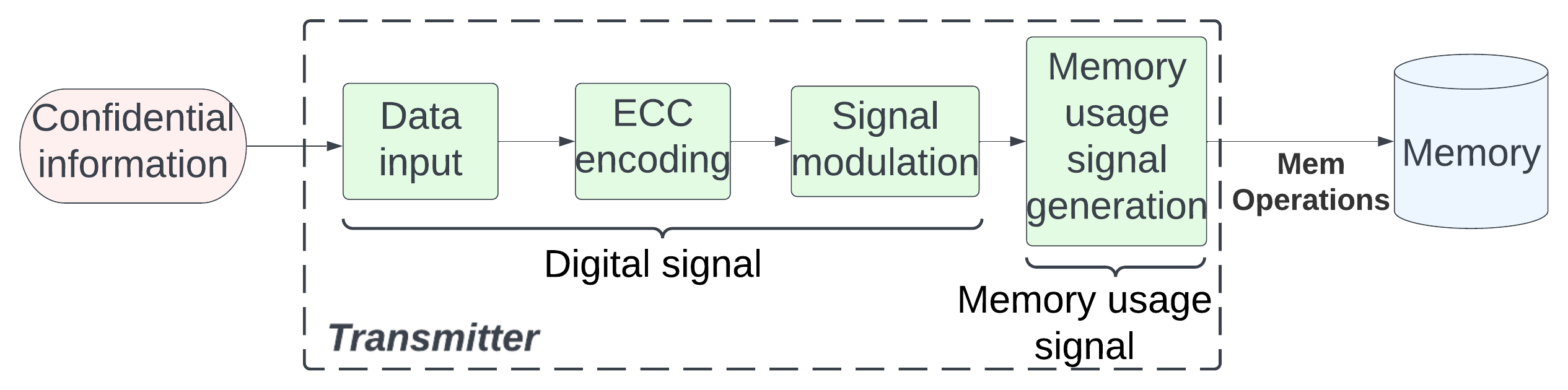}
  \caption{Overview of the transmitter module}
  \label{fig:mem_tx}
\end{figure}

As seen in \cref{fig:mem_tx}, the first step consists of converting these data to a binary, as well as separating them into blocks of 4 bits each. 
Then, the Hamming error correction code (ECC) is used in its 4-7 form ---taking a 4-bit input and converting it to a 7-bit ECC encoded output--- to minimize information loss due to the effects of temporal offsets caused by delays in the execution of memory operations, either by OS processes or other applications.

Furthermore, the data are also separated into packages of a fixed length to maintain integrity.
This way, if a package has more errors than the ECC can correct (a total of one bit flip in the 4-7 form), it is considered to be incorrect but may not affect the next packages to be transmitted.
We introduce a control sequence in front of each package in the form of a binary 1 appended to its most significant bit as a header, which indicates to the receiver that what follows is the package itself.
Therefore, the total length per package is 8 bits. 

Each bit is modulated by allocating, writing, and freeing memory,  using OOK (on-off keying), where the amplitude of the signal indicates the transmission of a `1' or a `0'.
A high amplitude value (i.e., high memory usage), which represents a `1', can be seen as a rise, and a 'low' constant flat value (i.e., no memory used) represents a logical `0' in the signal. 
This behavior corresponds to the communication line code RZ (return to zero). 
\Cref{eq:times_eq} shows how a logical `1' is represented, where $t_p$ is the time it takes to send a pulse, $t_h$ is the time when the signal rises (i.e., the total memory usage in the system goes up) and $t_l$ is the time where the signal returns to its low value (memory usage is similar to the one before).
On the other hand, a `0' would be a constant low signal of duration $t_p$ ($t_p = t_l$).

\begin{equation} \label{eq:times_eq}
    t_p = t_h + t_l,
\end{equation}

To build a pulse, we first need to increase the total system memory usage by copying a block of data to reserve it in memory and hold it for $t_h$, equivalent to $T/2$ where $T$ is the total pulse time. 
Then, the space is freed to lower the signal value once again. 
A sleep function is used to wait for the next pulse (low value in the remaining $T/2$), depending on whether a `1' or a `0' was sent.
This is done via software through a program written in C++.
The pseudocode algorithm for the transmitter module is shown in \cref{alg:mem:tx}.

\begin{algorithm}[ht]
\caption{Send Data through the Memory Covert Channel}
\label{alg:mem:tx}
\footnotesize
\KwIn{binaryString}
\KwOut{Transmitted Data}

\BlankLine
\textbf{Initialization:} \\
\quad BIT\_PULSE\_COUNT $\gets 2$ \\
\quad PACKAGE\_BIT\_COUNT $\gets 4$ \\
\quad SIZE\_BYTES $\gets 20 \times 1024 \times 1024$ \\
\quad REST\_MS $\gets 10$  \Comment{Assuming $t_h = t_l = 10ms$}
\BlankLine

\uIf{binaryString \% PACKAGE\_BIT\_COUNT $\neq 0$}{
    Append zeros to make binaryString divisible by PACKAGE\_BIT\_COUNT\;
}

encodedData = "" \;
\ForEach{package of PACKAGE\_BIT\_COUNT from binaryString}{
    encodedPackage = \texttt{hamming\_codec::encode}(package, PACKAGE\_BIT\_COUNT)\;
    encodedPackage = "1" + encodedPackage\;
    encodedData += encodedPackage\;
}

\BlankLine
\textbf{Sending the encoded data:} \\
\ForEach{bit in encodedData}{
    \If{bit = 1}{
        \For{$i = 1$ to BIT\_PULSE\_COUNT}{
            \For{$j = 1$ to SIZE\_BYTES}{
                Allocate $pBigArray[j] \gets 0xA$ \Comment{Write dummy data to the array}\\
                Copy $pBigArray$ to another array $pDestArray$\;
            }
            Free memory of $pBigArray$ and $pDestArray$\;
            Sleep for $REST\_MS$  milliseconds\;
        }
    }
    \Else{
        \For{$i = 1$ to BIT\_PULSE\_COUNT}{
            Sleep for $REST\_MS \times 2$ milliseconds\;
        }
    }
}
\BlankLine
\textbf{End of Transmission.}
\end{algorithm}

\subsubsection{\textbf{Receiver}}

\begin{figure} [ht]
    \centering
    \includegraphics[width=\linewidth]{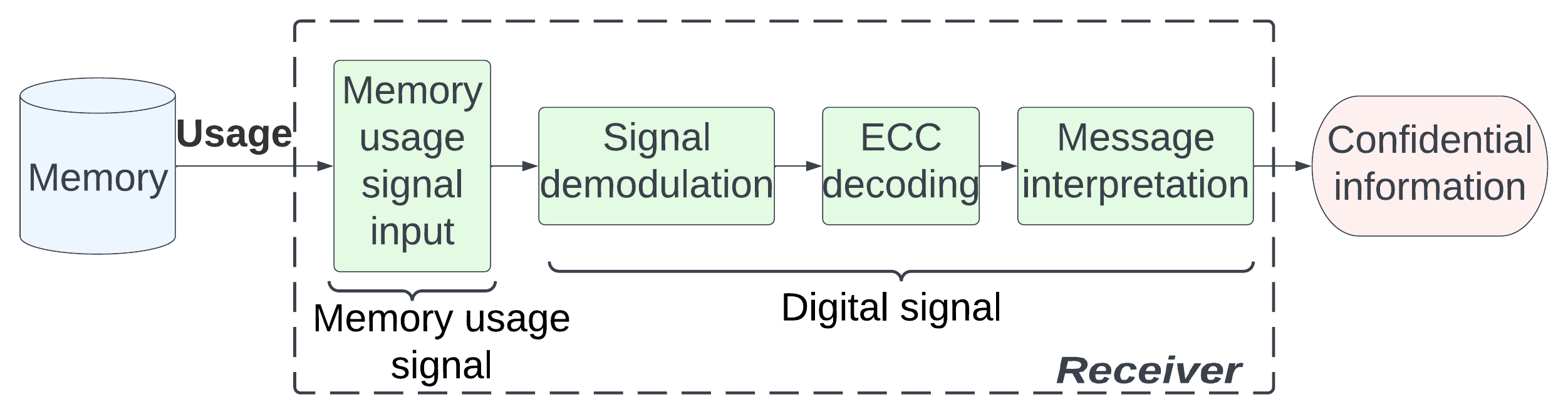}
    \caption{Overview of the receiver module}
    \label{fig:mem_rx}
\end{figure}

The receiver keeps track of the memory usage values throughout a period of time and with these data extracts the information that is being transmitted through the channel. 
The process implemented as the receiver module is depicted in \cref{fig:mem_rx}.
First, it samples the total memory usage by reading and parsing the `/proc/meminfo' file from the Linux OS, which by default does not require administrator privileges.
This value is calculated using \cref{eq:meminfo}, reading the corresponding fields of the file.
The sample rate is determined by the Nyquist theorem, which is set to at least double the channel frequency. 

\begin{equation} 
    mem_{used} = mem_{total} - mem_{free} - buffers - cache
    \label{eq:meminfo}
\end{equation}

Given that the modulated memory usage signal has a periodic behavior in which the pulses have a similar duration, we can group the receiver samples into sets of size $N$.
Each group would contain the representation of either a `1' or a `0'.

We implement a hybrid demodulation technique in this covert channel, working with the net total memory usage values over a period of time, as well as a frequency analysis to process and convert the channel signal to binary data. 
Because we know that the first bit (header) in a package is always `1', the receiver starts by looking for a pattern in the differences between the signal values that may correspond to the rise and fall of `1' (in the time domain).
Then it is very likely that from the first sample of the control bit, the next $8 \times N$ samples contain the transmitted package. 

The receiver then takes each set of samples and calculates its \ac{DFT} spectrum.
To minimize the effects of low-frequency interference in the signal, we include a 5th order Butterworth high-pass filter, to be applied before the binary translation.

If the set of samples housed a bit of `1', there will be a peak in amplitude tied to the frequency where the channel is transmitting, which is related to the frequency of the pulse used in the modulation process. 
Whereas, in a \ac{DFT} spectrum of a bit of `0', there will be no appreciable peaks in amplitude.
This classification is formalized in \cref{eq:windows}, where a `1' is tagged as such if its amplitude $A_i$ is greater than or equal to a defined threshold $A_0 (f_t)$, or as a `0' otherwise.

\begin{equation} \label{eq:windows}
  S_i=
  \begin{cases}
    1, & \text{if } A_i \geq A_0 (f_t) \\
    0, & \text{if } A_i < A_0 (f_t),
  \end{cases}
\end{equation}

Once the package is demodulated in binary form, what is left is to apply the reverse Hamming code to decode the message. 

In practice, this method achieves bit transmission errors less than 5\% in our transmission time, as shown in \cref{sec:eval:channel}, making it very reliable to send data across the two modules.

\section{Detection Technique and Countermeasures} \label{sec:detec_and__counter}

\subsection{ML-based Detector} \label{sec:detector}
In the environment of our novel covert channel, system memory utilization is intricately influenced by an array of external factors that encompass the operations, optimizations, and hardware characteristics of various running processes. 
The dynamic nature of these influences impacts the effectiveness of simple detection heuristics, such as fixed thresholds. 
Such approaches are inherently inadequate due to the absence of a universally applicable threshold capable of reliably predicting behaviors exhibited during both covert data transmission and routine system operations \cite{wang2023}.
Moreover, empirical evidence \cite{gonzalez23dotecca} has shown that regular applications can generate usage patterns in system resources that closely resemble those observed during covert transmissions. This behavioral similarity increases the challenge of distinguishing between normal activities and actual covert channels solely through simple heuristics, e.g., static threshold analysis.

Consequently, these observations advocate for the adoption of more sophisticated detection strategies beyond simplistic heuristics, such as threshold-based methods. 
To tackle this complexity, we proposed \textbf{a machine- learning-based detection technique}.
The rationale behind employing supervised learning methods for detection arises from the need to accommodate the continuous temporal changes in the memory utilization signal over time. 
Thus, the primary focus of our detection lies in developing robust models capable of learning and discriminating the behavioral patterns indicative of covert channel activities from normal system operations. This methodological approach enables the detector to proficiently analyze and classify intricate temporal signals, increasing the precision and dependability of covert channel detection in real-world  environments.
In order to show the complexity of detecting our new software-based covert channel and an initial point for a countermeasure, we perform a methodological exploration for a detection mechanism using different models from the ML domain.

\begin{figure} [ht]
    \centering
    \includegraphics[width=\linewidth]{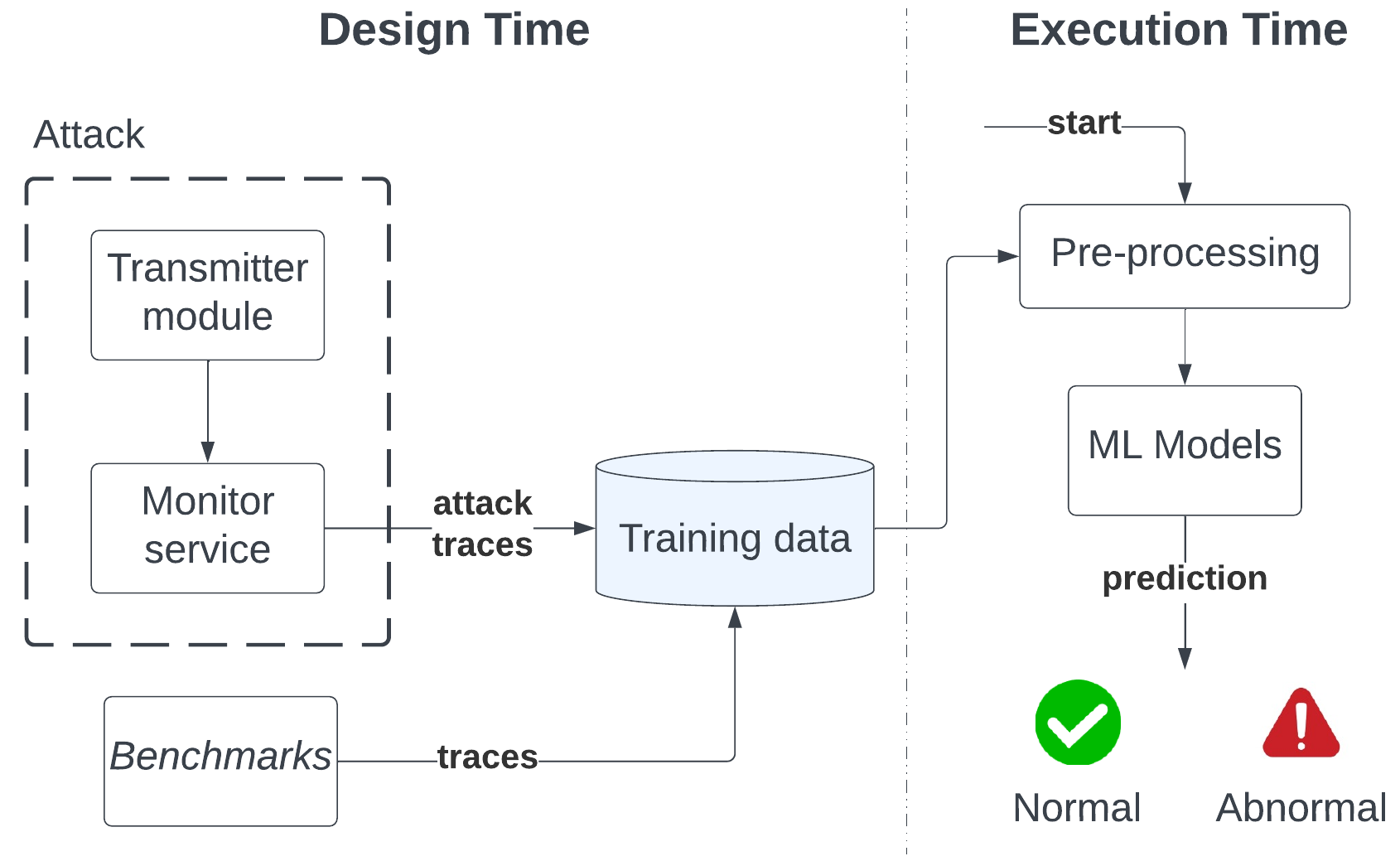}
    \caption{Overview of the detection technique}
    \label{fig:detect}
\end{figure}

An overview of the detection mechanism is given in \cref{fig:detect}.
At design time, we take samples of time-domain measurements from the memory usage in windows of 100 samples from our attack and the benchmark applications from the Phoronix suite~\cite{Phoronix}.
We form the data set this way to train different machine learning models, such as \acp{SVC}, \ac{KNN}, \acp{DT}, and \acp{NN}, using the scikit-learn Python library~\cite{scikit-learn}.

For the case of data extraction, a set consisting of 1,697,500 data points was used, divided evenly between 848,750 non-attack points and 848,750 attack points. These were extracted for the purpose of feeding the training models and subsequently being used for cross-validation.

Furthermore, the data set is transformed into a numerical array to facilitate efficient computations. 
Then a two-dimensional matrix is initialized on the basis of the dataset size and other predetermined parameters.
The data set undergoes normalization to scale the feature values between zero and one. 
A loop iterates through the matrix, assigning values based on specific conditions tied to the index and normalized data. 
The populated matrix is finally converted to a structured data frame for downstream analysis in the next steps.

Once the data are pre-processed, the machine learning models are trained and validated using the data set originating from a CSV file. 
This data set is explicitly labeled to indicate two classes: attacks marked as `1' and no attacks marked as `0'. 
The structured nature of this input allows for rigorous training of the machine learning algorithms, ensuring that they are well-equipped to differentiate between covert channel activity and benign memory usage patterns.

At run time, every 100ms, we take 100 samples from the system's memory utilization and apply the same preprocessing described above.
After this short task, the data are applied to the trained ML models to obtain the prediction of whether an attack is present in the system or not.

The system combines computational efficiency at run-time with fine-tuned hyperparameter settings to achieve high classification accuracy. 
The complete evaluation of the different models is presented in \cref{sec:eval_detection}.

\subsection{Countermeasure} \label{sec:counter}

Simple countermeasures to covert channels might seek to limit access to the sensor that the receiver accesses.
Although technically possible, limiting access to the memory usage information would definitely restrict other applications that rely heavily on this measurement.
As we show in \cref{sec:eval:channel}, our new covert channel is unaffected by background noise, which means that a more complex solution should be devised.
For this purpose, the proposed countermeasure consists of a focused noise generation application. 
Its inner workings and calibration are based on the transmitter module, but from a mitigation perspective.  
The assumptions to consider for the countermeasure are the following.
\begin{enumerate}
    \item There is confirmation from a detection mechanism (\cref{sec:detector}) of a present attack in the system.
    \item The predominant frequencies of the transmission are known or estimated, in order to target the generated noise to those specific frequencies.
    This channel frequency can be estimated by computing the \ac{DFT} over a short period of time (that is, a few seconds) while the attack is present.
\end{enumerate}
To generate the noise, pulses in memory usage are built similarly to the transmitter application, by allocating, writing, and freeing memory periodically at the same rate as the transmission frequency given by the detection technique. 
The goal of this countermeasure is not to remove the malicious application that is sending data, but rather to effectively cut the communication between transmitter and receiver modules ---thus disabling the covert channel--- from software-driven user space with the least manual configuration possible.
The evaluation of the countermeasure is presented in \cref{sec:eval_counter}.

\section{Evaluation}
\subsection{Evaluation platforms} \label{eval:platforms}
Our experiments were carried out on two different platforms: a general-purpose desktop PC and an embedded Raspberry Pi~4, with the characteristics shown in \cref{tab:platform-specs}. 
Both platforms intend to show the feasibility of the attack under different architectures while also using different transmission speeds.
Both devices implement the transmitter and receiver modules (covert channel) as well as the countermeasure. 
The transmitter and countermeasure are C++ applications, while the receiver module has a sampler monitor stage written in C++ and a data post-processing stage written in Python.

\begin{table}[ht]
\caption{Evaluation platforms}
\centering
\begin{tabular}{|c|cc|}
\hline
\multirow{2}{*}{\textbf{Characteristic}} & \multicolumn{2}{c|}{\textbf{Platform}}                               \\ \cline{2-3} 
                                         & \multicolumn{1}{c|}{\textbf{PC}}          & \textbf{Raspberry Pi 4b} \\ \hline
CPU                                      & \multicolumn{1}{c|}{Intel Core i9 10850K} & ARM Cortex-A72           \\ \hline
Cores and frequency                      & \multicolumn{1}{c|}{10 cores @ 3.6GHz}    & 4 cores @ 1.9GHz         \\ \hline
Memory & \multicolumn{1}{c|}{\begin{tabular}[c]{@{}c@{}}2x16GB DDR4 \\ @3600MHz\end{tabular}} & \begin{tabular}[c]{@{}c@{}}4GB LPDDR4 \\ @3200MHz\end{tabular} \\ \hline
Operating system                         & \multicolumn{1}{c|}{Ubuntu 22.04}         & Linux (buildroot)        \\ \hline
\end{tabular}
\label{tab:platform-specs}
\end{table}

\subsection{Covert channel metrics} \label{sec:eval:channel}

\begin{table}[htp]
\caption{Covert channel evaluation results}
\begin{center}
\label{tab:metrics}
\begin{tabular}{|c|c|c|c|c|}
\hline
\textbf{Platform} &
  \textbf{Bits sent} &
  \textbf{\begin{tabular}[c]{@{}c@{}}Transmission\\ speed\end{tabular}} &
  \textbf{\begin{tabular}[c]{@{}c@{}}BER \\ (\%)\end{tabular}} &
  \textbf{\begin{tabular}[c]{@{}c@{}}PER \\ (\%)\end{tabular}} \\ \hline
PC &
  337,952 &
  6.5 &
  0.32 &
  0.71 \\ \hline
RPi~4 &
  4096 &
  125 &
  2.23 &
  15.7 \\ \hline
\end{tabular}
\end{center}
\end{table}

To evaluate the effectiveness of the attack, we devised two experiments using the evaluation platforms described above. 
The summary of these experiments can be seen in \cref{tab:metrics}.
First, as the main evaluation for our channel, we employ the general computing PC platform.
In this experiment, we send a total of 337,952 bits through the channel.
This corresponds to 42,244 packages, each with a length of 8 bits, which include a mixture of ASCII-encoded messages, random bit strings, and images with different types of encoding schemes. 
Our new software-driven covert channel attack operates at a net speed of 6.25 bits per second (bps) with a frequency of 25Hz on this platform.
The memory block that the transmitter reserves and releases to build the pulse is 20MB in size.
Samples of total memory used by the computer, calculating it like shown in \cref{eq:meminfo}, are logged every 1 ms.

As depicted, the average bit error rate (BER) and the packet error rate (PER) were below 0. 5\% and 1\%, respectively, with many individual messages achieving perfect transmission (0\% error rates). 
This indicates a high reliability of this new covert attack for the leaking of confidential user data in a real setting at a reasonable transmission speed. 

\begin{figure} [ht]
    \centering
    \includegraphics[width=\linewidth]{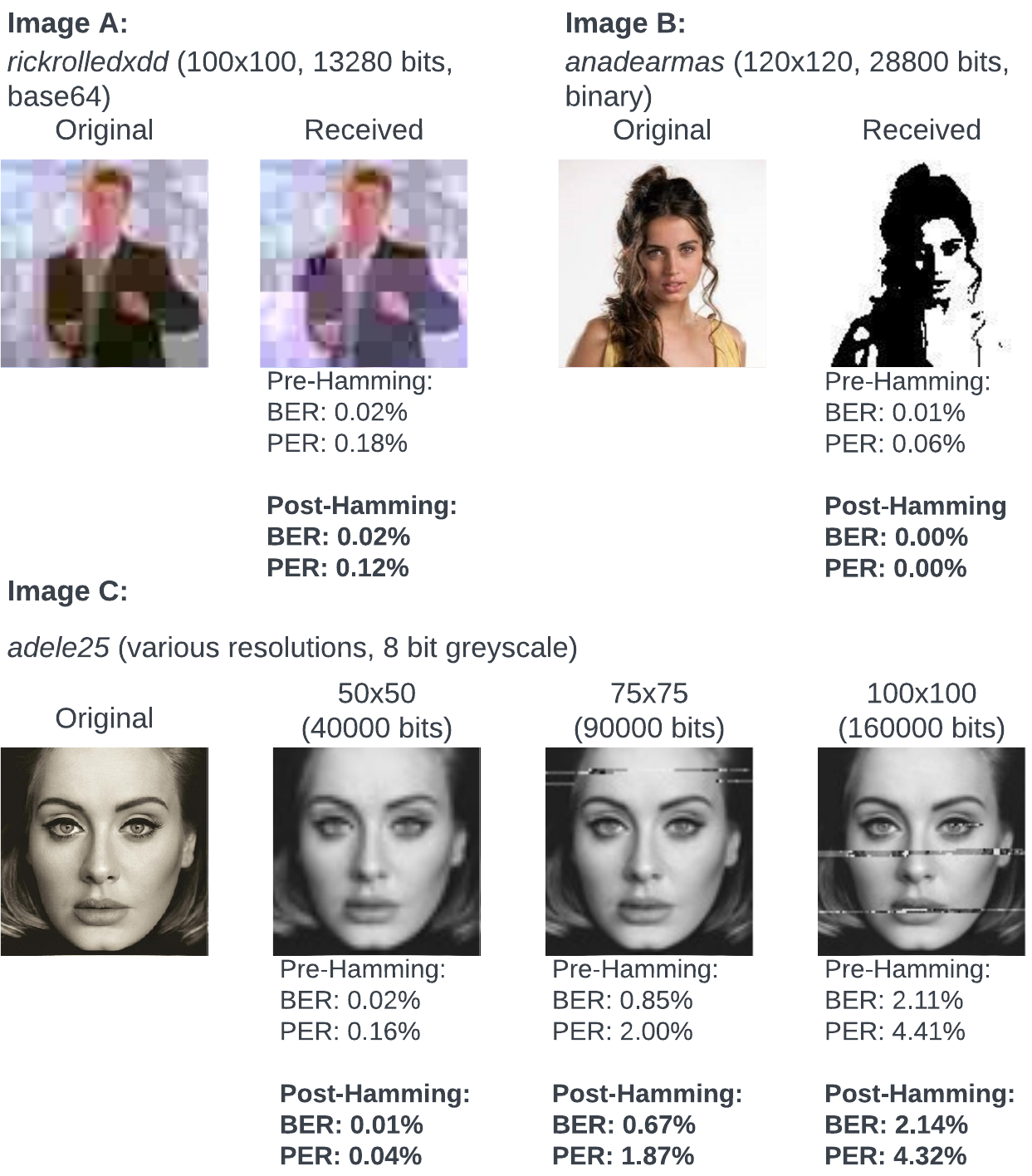}
    \caption{Visual demonstration of the effectiveness of the software-driven memory usage covert channel}
    \label{fig:visual}
\end{figure}

To visually demonstrate how secret information can be sent through the covert channel, we tested encoded images in various formats, as seen in \cref{fig:visual}.
This shows that even if the error rate may not be 0\% in all cases, what these images represent and their details can be easily interpreted by a potential malicious agent.

As a second experiment, to demonstrate how the attack can be implemented in a hardware-independent way, we use the embedded Raspberry Pi~4 as the target.
Here, we transmit 4096 bits at a much faster channel transmission frequency to show the versatility of the attack.
Although the packet error rate is higher due to increased speed, the bit error rate remains very low.
More importantly, no changes were required from the original PC implementation, which shows how this new software-based covert channel does not depend on specific hardware components.

\subsection{Covert channel under background noise} \label{sec:eval:noise}
In order to evaluate the effect of other background applications (and O.S) on the channel, we devise an experiment in which we run applications benchmark from the Phoronix Suite alongside our covert channel in the PC platform.
The results of this experiment can be seen in Table \cref{tab:noise}.
As shown, even when the different applications execute producing multiple background memory accesses, the covert channel is able to effectively communicate all the packets with a very low error rate of less than $5\%$.
These results confirm the robustness of the channel even when transmitted under a realistic noise background from the operating system and other applications.

\begin{table}[ht]
\caption{Effect of background application noise in the memory-based covert channel}
\label{tab:noise}
\centering
\begin{tabular}{|l|c|c|c|}
\hline
\multicolumn{1}{|c|}{\textbf{Test application}} &
  \textbf{\begin{tabular}[c]{@{}c@{}}Packets \\ sent\end{tabular}} &
  \textbf{\begin{tabular}[c]{@{}c@{}}BER \\ (\%)\end{tabular}} &
  \textbf{\begin{tabular}[c]{@{}c@{}}PER \\ (\%)\end{tabular}} \\ \hline
lighthouse\_chorus\_cachebench & 222 & 0.11 & 0.90 \\ \hline
lighthouse\_vp (1080p video)   & 56  & 0.00 & 0.00 \\ \hline
lighthouse\_vp4k (4k video )   & 56  & 2.68 & 3.57 \\ \hline
lighthouse\_vp4k\_2            & 56  & 0.45 & 3.57 \\ \hline
lighthouse\_vp4k\_3            & 56  & 0.00 & 0.00 \\ \hline
sg\_game (Left 4 Dead 2)       & 76  & 0.00 & 0.00 \\ \hline
lh\_game (Left 4 Dead 2)       & 280 & 0.80 & 2.86 \\ \hline
lh\_br\_game (Left 4 Dead 2)   & 316 & 1.11 & 4.43 \\ \hline
\multicolumn{1}{|c|}{\textbf{TOTAL / AVERAGE \%}} &
  \textbf{1118} &
  \textbf{0.64} &
  \textbf{1.92} \\ \hline
\end{tabular}
\end{table}

\subsection{Real use case: VM to host communication through MeMoir in a Hyper-V environment}

To evaluate the effectiveness of our memory-based covert channel attack in a real scenario, we conducted an experiment utilizing a virtualized environment with Windows Subsystem for Linux 2 (WSL 2) running on a Windows 11 host, on the PC platform described in \cref{tab:platform-specs}

WSL 2, a compatibility layer for running Linux binaries on Windows, leverages a full Linux kernel within a lightweight virtual machine managed by Microsoft's Hyper-V \cite{MicrosoftHyperV}. 
Hyper-V's robust virtualization infrastructure enables WSL 2 to dynamically allocate and deallocate memory based on the workload within the virtual machine. 
This dynamic memory management feature is crucial to our covert channel, as it allows the memory usage of the WSL 2 VM to fluctuate in response to specific actions taken by processes inside the Linux environment. By monitoring these fluctuations in memory usage from the Windows 11 host, we can infer the transmitted information.

\begin{figure}[ht]
    \centering
    \includegraphics[width=0.8\linewidth]{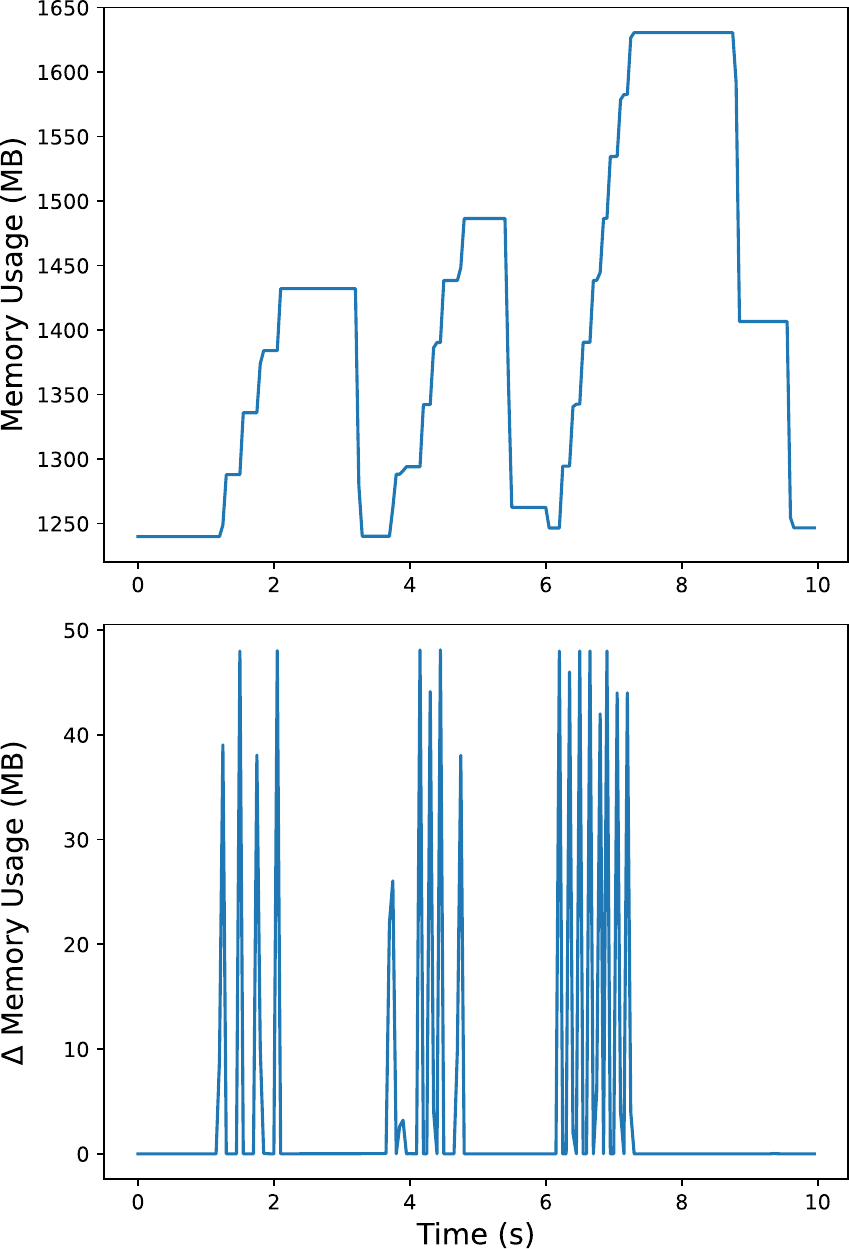}
    \caption{Memory usage signal (top) and processed memory $\Delta$ (bottom) measured from the host machine's \textit{vmmemWSL} process when transmitting three 8-bit packets: b'10101010 (0xAA), b'10011101 (0x9D) and b'11111111 (0xFF), at a transmission rate of $5$~bps.}
    \label{fig:memwsl}
\end{figure}

In our experimental setup for this real use case, the memory-based covert channel operated by carefully orchestrating memory usage patterns inside the WSL 2 instance, employing a nonreturn-to-zero encoding, where an increase ($\Delta$) in the memory usage corresponds to the enconding of a bit of a `1', while a constant memory usage corresponds to an enconding of a `0'. 
These patterns were designed to create detectable changes in the memory metrics observed from the host system. 
The memory usage of the virtualized environment is done via the \textit{vmmemWSL} process in the host system, which is responsible for managing the WSL 2 virtual machine. 
This monitoring was performed at a high frequency (that is, 50 Hz) to capture subtle changes in memory allocation and deallocation.
To better visualize the applied encoding,  \cref{fig:memwsl} shows the memory usage of the VM for three packets at a transmission rate of 5 bits per second (bps).
As it can be further extracted from the figure, a threshold  ($\delta$) of about 25MB on the $\Delta$ memory usage signal, is sufficient to differentiate between `1' and `0' on the receiver's side.

\begin{table}[h]
\caption{Metrics for VM-to-host covert channel communication}
\label{tab:hvm}
\centering
\begin{tabular}{|c|c|c|c|c|}
\hline
\textbf{Bits sent} & \textbf{Packets} & \textbf{Bit rate (bps)} & \textbf{BER (\%)} & \textbf{PER (\%)} \\ \hline
8,000 & 1,000 & 5 & 1.05 & 3.70 \\ \hline
\end{tabular}
\end{table}

In order to further validate the VM-to-host covert channel, we performed an experiment where we sent 1,000 8-bit packets from the VM to the host environment, at a local transmission rate of 5 bits per second (bps). 
The results of this experiment can be seen in \cref{tab:hvm}.
As depicted, the very low bit and error rates from this experiment (i.e., less than 5\%) highlight the feasibility of our memory-based covert channel under a realistic use case, by leveraging dynamic memory allocation features inherent in WSL 2 and Microsotf's Hyper-V to facilitate covert communication channels.

\subsection{Covert channel detection} 
\label{sec:eval_detection}
In order to evaluate the covert channel detection mechanism described in \cref{sec:detector}, we implement multiple machine learning algorithms and measure  their effectiveness in detecting attacks.
Each algorithm, including Support Vector Classifiers (\ac{SVC}), k-Nearest Neighbors (\ac{KNN}), Decision Trees (\ac{DT}), and Neural Networks (\ac{NN}), undergoes thorough testing on performance metrics such as accuracy, false negatives (FN), false positives (FP), and execution time using the PC platform from \cref{eval:platforms}. 
The goal of these experiments is to determine which method offers the most reliable and efficient solution for detecting malicious activities in the system.

\ac{SVC} algorithms exhibit varying performance metrics as depicted in \cref{tab:svc-performance}. 
Among them, the standard \ac{SVC} algorithm has the highest accuracy of \(89.48\%\) but takes significant time to execute.
Linear\ac{SVC} and Nu\ac{SVC} are slightly less accurate but have different trade-offs in terms of false negatives and false positives.

\begin{table}[ht]
\caption{Performance Metrics for \ac{SVC} Algorithms}
\centering
\setlength{\tabcolsep}{5pt} 
\begin{tabular}{|c|c|c|c|c|}
\hline
\textbf{Alg.} & \textbf{Acc. (\%)} & \textbf{FN (\%)} & \textbf{FP (\%)} & \textbf{Time (ms)} \\ \hline
\ac{SVC}           & 89.48              & 12.33            & 6.58             & 5020.00 \\ \hline
Linear\ac{SVC}     & 87.72              & 8.46             & 13.24            & 3370.00 \\ \hline
Nu\ac{SVC}         & 88.01              & 9.54             & 11.81            & 16700.00 \\ \hline
\end{tabular}
\label{tab:svc-performance}
\end{table}

\ac{KNN} algorithms (\cref{tab:knn-performance}) show superior accuracy, with \ac{KNN} 10 being the best configuration at \(94.85\%\). 
However, the accuracy starts to drop as the number of neighbors increases, although the execution time is relatively low.
It is worth mentioning that false positives are extremely low or even zero across different configurations.

\begin{table}[ht] 
\caption{Performance metrics for \ac{KNN} algorithms with different neighbors}
\centering
\setlength{\tabcolsep}{4pt} 
\begin{tabular}{|c|c|c|c|c|}
\hline
\textbf{Alg. (Neighbors)} & \textbf{Acc. (\%)} & \textbf{FN (\%)} & \textbf{FP (\%)} & \textbf{Time (ms)} \\ \hline
\textbf{KNN 10 } & \textbf{94.85} & \textbf{9.44}  & \textbf{0.0}  & \textbf{14.00} \\ \hline
KNN 15  & 93.81 & 11.12 & 0.0  & 18.90 \\ \hline
KNN 20  & 92.72 & 12.83 & 0.0  & 13.40 \\ \hline
KNN 25  & 91.9  & 13.68 & 0.52 & 15.20 \\ \hline
\end{tabular}
\label{tab:knn-performance}
\end{table}

The \acp{DT} show an increasing trend in accuracy with the depth of the tree, as shown in \cref{tab:decision-tree-performance}. 
Remarkably, DT~7 has an impressive accuracy of $97.64\%$, albeit at a slightly higher computational cost.
False positives are also minimal, making it a strong choice.

\begin{table}[ht]
\caption{Performance metrics for \acp{DT} with different max depths}
\centering
\setlength{\tabcolsep}{5pt} 
\begin{tabular}{|c|c|c|c|c|}
\hline
\textbf{Alg. (Depth)} & \textbf{Acc. (\%)} & \textbf{FN (\%)} & \textbf{FP (\%)} & \textbf{Time (ms)} \\ \hline
DT 1  & 90.04 & 12.83 & 5.03  & 119.00 \\ \hline
DT 3  & 91.05 & 8.16  & 8.28  & 330.00 \\ \hline
DT 5  & 94.11 & 6.73  & 4.40  & 470.00 \\ \hline
\textbf{DT 7}  & \textbf{97.64} & \textbf{4.50} & \textbf{0.06}  & \textbf{581.00} \\ \hline
\end{tabular}
\label{tab:decision-tree-performance}
\end{table}

Finally, \acp{NN}, shown in \cref{tab:neural-network-performance}, have a max accuracy of $90.87\%$ but require significantly moderate execution times. 
The architecture of our \ac{NN} is designed with an input layer that is dynamically configured to match the shape of the training data, featuring configurations for 16 and 8 neurons.
This is followed by the first hidden layer with 8/4 neurons, a second hidden layer with 4/2 neurons, and finally culminating in a single-neuron output layer.
Each hidden layer performs specific transformations that contribute to the model's specialization in binary classification.
The table shows the metrics from four selected architectures with different activation functions.
Generally, models also have a relatively high rate of false negatives, which makes them less reliable for detecting certain types of attacks.

\begin{table}[ht]
\caption{Performance metrics for \acp{NN} with different activation functions}
\label{tab:nn}
\centering
\setlength{\tabcolsep}{5pt} 
\begin{tabular}{|c|c|c|c|c|}
\hline
\textbf{Alg. (Arch.)} & \textbf{Acc. (\%)} & \textbf{FN (\%)} & \textbf{FP (\%)} & \textbf{Time (ms)} \\ \hline
NN (16-8-4-1, ReLU) & 90.54 & 12.15 & 4.93 & 799.00 \\ \hline
NN (16-8-4-1, tanh) & 15.99 & 49.11 & 41.80 & 476.00 \\ \hline
NN (8-4-2-1, Sigmoid) & 89.01 & 11.96 & 7.79 & 465.32 \\ \hline
NN (8-4-4-1, SeLU) & 87.42 & 9.20 & 13.02 & 399.92 \\ \hline 
\end{tabular}
\label{tab:neural-network-performance}
\end{table}

In general, the optimal algorithm depends on the specific needs and constraints of the system in question. 
DT with a maximum depth of 7 excels in accuracy and false negative and positive rates, but it may not be suitable for real-time or large-volume data processing due to its higher inference time. 
On the other hand, \ac{KNN} with 10 neighbors offers a balance between good accuracy and faster inference, although at the expense of a slightly higher false negative rate. 
Ultimately, both models obtained high detection accuracy with acceptable false positive and negative rates and execution times, showing that ML-based detection techniques are feasible for this type of attack.

\subsection{Countermeasure evaluation} \label{sec:eval_counter}
To evaluate the performance of our countermeasure, we implement the noise-based technique described in \cref{sec:counter}.
We evaluated how effective the technique was at blocking the attack in both of our evaluation platforms.
The results of this experiment are shown in \cref{tab:counter_ber}.
\begin{table}[htp]
\caption{Bit and packet error of the cover channel rates under the proposed countermeasure}
\label{tab:counter_ber}
\begin{center}
\begin{tabular}{|c|c|c|}
\hline
\textbf{Platform} & \textbf{BER (\%)} & \textbf{PER (\%)} \\ \hline
PC                & 53.21             & 97.58             \\ \hline
Raspberry Pi 4            & 46.03             & 100               \\ \hline
\end{tabular}
\end{center}
\end{table}

As noted, the proposed countermeasure was successful in cutting communication through the channel on both platforms. 
While active, BER was around 50\%, while PER reached over 97\% for both platforms, even after Hamming ECC decoding. 
BER is expected to be around 50\% because we use a binary system to encode the data (1/2 chance of randomly selecting `0' or `1').
In any case, no meaningful data is received.
This is confirmed by the fact that the vast majority of packets show errors while the countermeasure is active.

\begin{table}
\vspace{-5px}
\caption{Average power for the countermeasure and benchmarks compared to an idle state}
\label{tab:power}
\centering
\begin{tabular}{|c|cc|}
\hline
\multirow{2}{*}{\textbf{System state}} & \multicolumn{2}{c|}{\textbf{Average power (W)}} \\ \cline{2-3} 
                                       & \multicolumn{1}{c|}{CPU cores}      & DRAM      \\ \hline
Idle                                   & \multicolumn{1}{c|}{4.77}           & 0.47      \\ \hline
Countermeasure                         & \multicolumn{1}{c|}{31.28}          & 1.00      \\ \hline
Benchmarks                             & \multicolumn{1}{c|}{33.18}          & 0.91      \\ \hline
\end{tabular}
\label{tab:power-consumption}
\end{table}

Furthermore, we performed a system power consumption analysis for different states for the PC platform as seen in \cref{tab:power}.
We used the power estimation feature RAPL (Running Average Power Limit)~\cite{Rapl2010}, included in most modern Intel processors, to estimate the current power draw from its CPU cores and the system's DRAM via a monitoring application measuring at a sample rate of 1 ms. 
The benchmarks were run through the Phoronix suite~\cite{Phoronix}, and feature the tests CacheBench (memory and cache bandwidth performance stress), 4K video playback, LibreOffice, and the video games Left 4 Dead 2 and Counter-Strike: Global Offensive running at 4K at the maximum video settings for this machine. 
From the experiment, we observe that while the countermeasure is active, the system consumes on average slightly less power than the benchmark applications.
This means that while there is an associated cost for the countermeasure, its application is not more computationally expensive than a simple benchmark application.

\section{Conclusion}
In this paper, we have proposed a new software-driven covert channel using the system's memory usage as the medium for the attack. Our results have shown how this attack is practical under two different platforms, achieving high transmission rates with low errors. 
Moreover, we have presented a real-case scenario for our attack, where our novel memory-based covert channel allowed us to communicate information from a Hyper-V virtualized environment to a Windows host system with transmission errors less than 5\%. 
In addition, we have explored different alternatives for machine learning-based detection techniques for this new attack. Our exhaustive evaluation has shown how different ML models are able to quickly detect attacks with an accuracy of over 95\% and low false positives and false negatives.
Finally, we have implemented a noise-based countermeasure for the covert channel, which is able to mitigate the attack at the cost of additional power comparable to an average benchmark application.

\bibliographystyle{IEEEtran}
\bibliography{refs.bib}
\end{document}

%% file: acronyms.tex
\usepackage{acro}
\DeclareAcronym{ML}{short=ML,long=machine learning}
\DeclareAcronym{DFT}{short=DFT,long=Discrete Fourier Transform}

\DeclareAcronym{SVC}{short=SVC,long=Support Vector Machine Classificator}

\DeclareAcronym{KNN}{short=KNN, long=K-Nearest Neighbors}
\DeclareAcronym{NN}{short=NN, long=Neural Network}
\DeclareAcronym{DT}{short=DT, long=Decision Tree}